\documentclass[doublecol]{epl2}
\usepackage{amssymb,amsmath,graphicx,color,rotating,subfigure,url,multirow}

\title{Online-offline activities and game-playing behaviors of avatars in a
massive multiplayer online role-playing game}
\shorttitle{Online-offline activities and game-playing behaviors of avatars} 

\author{Zhi-Qiang Jiang\inst{1,2,3} \and Wei-Xing Zhou\inst{1,2,3,4}\footnote{e-mail: wxzhou@ecust.edu.cn} \and Qun-Zhao Tan \inst{5}}
\shortauthor{Z.-Q. Jiang \etal}

\institute{
  \inst{1} School of Business, East China University of Science and Technology, Shanghai 200237\\
  \inst{2} School of Science, East China University of Science and Technology, Shanghai 200237\\
  \inst{3} Research Center for Econophysics, East China University of Science and Technology, Shanghai 200237\\
  \inst{4} Research Center on Fictitious Economics and Data Science, Chinese Academy of Sciences, Beijing 100190\\
  \inst{5} Shanda Interactive Entertainment Ltd, Shanghai 201203
}


\abstract{Massive multiplayer online role-playing games (MMORPGs)
are very popular in China, which provides a potential platform for
scientific research. We study the online-offline activities of
avatars in an MMORPG to understand their game-playing behavior. The
statistical analysis unveils that the active avatars can be
classified into three types. The avatars of the first type are owned
by game cheaters who go online and offline in preset time intervals
with the online duration distributions dominated by pulses. The
second type of avatars is characterized by a Weibull distribution in
the online durations, which is confirmed by statistical tests. The
distributions of online durations of the remaining individual
avatars differ from the above two types and cannot be described by a
simple form. These findings have potential applications in the game
industry.}

\begin{document}

\maketitle

\section{Introduction}

According to the Statistical Reports on the Internet Development in
China released by China Internet Network Information Center, the
past twelve years have witnessed a sharp increase in the number of
Chinese netizens from 0.63 million on 31 October 1997 to 338 million
on 16 July 2009. Till June 2009, the size of netizens playing
massive multiplayer online games (MMOGs) is 78.55 million. The MMOGs
in mainland China include two types, {\em{i.e.}}, the massive
multiplayer online role-playing game (MMORPG) and the large-scale
casual game, both having about 49 million users. An MMOG is an
online virtual world, where avatars can live and interact with one
another in a somewhat realistic manner. The huge number of users in
MMOGs has raised many open academic problems and attracted vast
interest of academics from diverse angles of view, especially since
the pioneering work done by Edward Castronova, who traveled in a
virtual world called ``Norrath'' and performed a preliminary
analysis of its economy \cite{Castronova-2001-WP}. Particularly,
virtual worlds have great potential for research in social,
behavioral, and economic sciences \cite{Bainbridge-2007-Science}.

The outbreak of SARS virus in 2003 and the recent globally spread
swine flu forces scientists to understand the epidemics of
infectious diseases. A lot of epidemic models have been proposed
\cite{Colizza-Barrat-Barthelemy-Vespignani-2006-BMB}. Although there
are several exceptions
\cite{Brokmann-Hufnagel-Geisel-2006-Nature,Wang-Gonzalez-Hidalgo-Barabasi-2009-Science},
the limited availability of empirical data of human mobility remains
a crucial challenge
\cite{Balcan-Colizza-Goncalves-Hu-Ramasco-Vespignani-2009-XXX}. To
partly overcome this difficulty, we can design a kind of virus in a
virtual world and let it spread to investigate its epidemics. For
other applications, we can design some economic games in a virtual
world to study the formation of human cooperation (indeed, numerical
experiments have been done \cite{Grabowski-Kosinski-2008-APPA}), and
we can record the economic behaviors of avatars to understand the
evolution of wealth distribution. There are also efforts in the
field of computational social sciences from a complex network
perspective
\cite{Grabowski-2007-PA,Grabowski-Kruszewska-2007-IJMPC,Grabowski-Kruszewska-Kosinski-2008-EPJB,Grabowski-Kosinski-2008-APPB,Grabowski-Kruszewska-Kosinski-2008-PRE,Grabowski-2009-PA}.
In addition to its scientific potentials, virtual worlds could act
as nice places for real social activities, such as marketing
\cite{Matsuda-2003-Presence,Castronova-2005-HBR,Hemp-2006-HBR}, and
provide opportunities for players to make real money
\cite{Papagiannidis-Bourlakis-Li-2008-TFSC}.

In this Letter, we investigate the online-offline activities and
game-playing behaviors of the avatars inhabiting a server of a
massive multiplayer online role-playing game operated by Shanghai
Shanda Interactive Entertainment Ltd, which is the leader of China's
MMORPG industry and runs dozens of online games. We will show that
the statistical properties of the online-offline activities of
individual avatars allow us to classify avatars and identify game
cheaters.

\section{Description and preprocessing of the data}

Our data are online-offline logs recorded during the time period
from 1 September 2007 to 31 October 2007 of an MMORPG server run by
Shanda Interactive Entertainment Ltd. There is one log file for each
day. Each entry contains three pieces of information: the masked
avatar ID, its login time, and its logout time. The resolution of
the time stamps is 1 second. During the recording time period, there
were 19843 avatars who entered the game. For security sake, the true
avatar IDs have been encrypted into numbers from 1 to 19843.

\begin{figure}[htp]
\centering
\includegraphics[width=6.5cm]{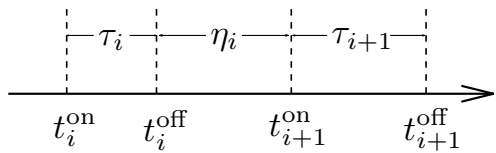}
\caption{\label{Fig:VW:tau:Definition} Schematic chart of game
sessions for an individual avatar and the definition of online
durations.}
\end{figure}

An entry is written to the log file when an avatar goes offline.
Therefore, the entries in a log file are arranged according to an
increasing order of logoff moments. For each avatar, we collect all
the associated entries, whose login and logoff times form a
two-dimensional array $E_{m\times2}$
\begin{equation}
 E_{m\times2} =
 \left(
  \begin{array}{cc}
    t_1^{\rm{on}} & t_1^{\rm{off}} \\
    \vdots & \vdots \\
    t_i^{\rm{on}} & t_i^{\rm{off}} \\
    \vdots & \vdots \\
    t_m^{\rm{on}} & t_m^{\rm{off}} \\
  \end{array}
\right),
\end{equation}
where $t_i^{\rm{on}}$ and $t_i^{\rm{off}}$ are the logon and logoff
times of the $i$-th game-playing session of the avatar during the
time period from 1 September 2007 to 31 October 2007. In the usual
situation, we have
\begin{equation}
 \cdots < t_i^{\rm{on}} <
 t_i^{\rm{off}} < t_{i+1}^{\rm{on}} < \cdots,
 \label{Eq:t1t2t3}
\end{equation}
which is illustrated in fig.~\ref{Fig:VW:tau:Definition}.

We can calculate the time interval $\tau_i$ between the logon time
$t_i^{\rm{on}}$ and logoff time $t_i^{\rm{off}}$ of the $i$-th game
session that an avatar played during the time period under
investigation,
\begin{equation}
\tau_i = t_i^{\rm{off}} - t_i^{\rm{on}},
 \label{Eq:gd}
\end{equation}
which is termed as {\em{online duration}} of the $i$-th game
session. We can also calculate the {\em{offline duration}} between
two successive game sessions of a same avatar as follows
\begin{equation}
\eta_{i} = t_{i+1}^{\rm{on}} - t_{i}^{\rm{off}},
 \label{Eq:wt}
\end{equation}
which measures how long it takes for an avatar to logon the game
again after he/she exits the game.

Assume that the sequence sizes of online and offline durations of
avatar $j$ are $n^{\rm{on}}_j$ and $n^{\rm{off}}_j$, respectively.
Since each $t_i^{\rm{on}}$ is followed by $t_i^{\rm{off}}$, we have
\begin{equation}
  n^{\rm{on}}_j = n^{\rm{off}}_j + 1.
  \label{Eq:n:on:off}
\end{equation}
Defining that $N^{\rm{on}} = \sum_{j=1}^{19843} n^{\rm{on}}_j$ and
$N^{\rm{off}} = \sum_{j=1}^{19843} n^{\rm{off}}_j$, it follows
immediately that
\begin{equation}
  N^{\rm{on}} = N^{\rm{off}}+19843.
  \label{Eq:N:on:off}
\end{equation}
We have calculated the online and offline duration sequences of all
the 19843 avatars and find that $N^{\rm{on}}=14,393,332$ and
$N^{\rm{off}} = 14,373,489$, which is consistent with
Eq.~(\ref{Eq:N:on:off}). On average, each avatar plays about 12
sessions each day.

Preprocessing the data is necessary. We find that there are 41,845
offline durations (about 0.3\% of the total sample) that are
negative, which can be attributed to recording errors introduced by
the system. There are also 1,221,811 offline durations (about 8.5\%
of the total sample) that equal to zero. The observation of $\eta =
0$ is nothing but a consequence of the data recording rule that the
log file will record the action that one avatar enters map B from
map A as an offline-online activity. For the above cases, we adopt
the strategy of removing the offline entry by merging the two
entries $\{t_i^{\rm{on}}, t_i^{\rm{off}}\}$ and
$\{t_{i+1}^{\rm{on}}, t_{i+1}^{\rm{off}}\}$ into one
$\{t_i^{\rm{on}}, t_{i+1}^{\rm{off}}\}$. It is possible that the
offline duration associated with an inter-map transfer of an avatar
is greater than 0 if there is a heavy network traffic. For the
online durations, all $\tau$ values are nonnegative and there are
52,442 online durations (about 0.4\% of the total sample) that are
equal to 0. The online durations with $\tau=0$ are excluded from
further analysis.

\section{Collective behaviors}

The instant number of online avatars per second can be constructed
according to the online-offline data, whose statistical properties
have been investigated \cite{Jiang-Ren-Gu-Tan-Zhou-2009-PA}. It was
found that the online avatar number exhibits one-day periodic
behavior and clear intraday pattern, the fluctuation distribution of
the online avatar numbers has a leptokurtic non-Gaussian shape with
power-law tails, the increments of online avatar numbers after
removing the intraday pattern are uncorrelated and the associated
absolute values have long-term correlation, and both time series
exhibit multifractal nature \cite{Jiang-Ren-Gu-Tan-Zhou-2009-PA}.
These properties are relevant to the traffic of the server and the
profit of the MMORPG company.

In this section, we will investigate the collective behaviors of
individual avatars based on their gaming activities. Three
quantities are studied. For each player, we define two quantities,
one is total online times $m$ and the other is total online session
duration $T$, and then take the whole population as a sample to make
a description of the collective activities.

\subsection{Distribution of the number of gaming sessions of
individual avatars}

\begin{figure}[htp]
\centering
\includegraphics[width=6.5cm]{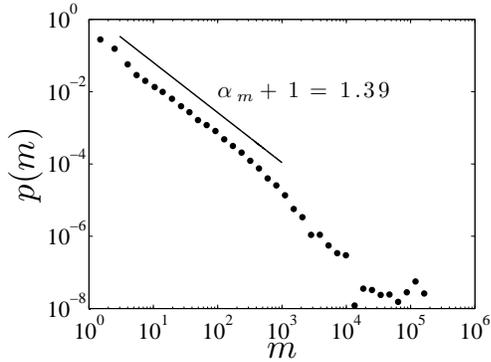}
\caption{\label{Fig:VW:PDF:m} Empirical probability density function
$p(m)$ of the number of sessions $m$ of 19843 individual avatars.}
\end{figure}

For each avatar, we count the number $m$ of gaming sessions that
he/she played during the two-month time period under investigation.
The sequence has 19843 data points. The empirical probability
density function $p(m)$ of individual gaming session number $m$ is
illustrated in fig.~\ref{Fig:VW:PDF:m}. One can observe that there
is a power-law behavior between $p(m)$ and $m$:
\begin{equation}
 p(m)\approx m^{-(\alpha_m+1)}, ~~{\rm{for}}~~m\geqslant{m_{\min}},
 \label{Eq:pdf:m}
\end{equation}
where the power-law exponent can be approximatively obtained by the
following equation based on the maximal likelihood estimation
\cite{Clauset-Shalizi-Newman-2009-SIAMR},
\begin{equation}
 \alpha_m = N_m \sum_{j=1}^{N_m} \ln
 \frac{m_j}{m_{\min}-0.5},
 \label{Eq:gamma}
\end{equation}
where $N_m$ is the number of $m$ that are no less than $m_{\min}$.
By setting $m_{\min} = 1$, Eq.~(\ref{Eq:gamma}) gives that the tail
exponent $\alpha_m = 0.39$. The Kolmogorov-Smirnov test confirms
that the distribution can model the data with high statistical
significance.

\begin{figure}[htp]
\centering
\includegraphics[width=6.5cm]{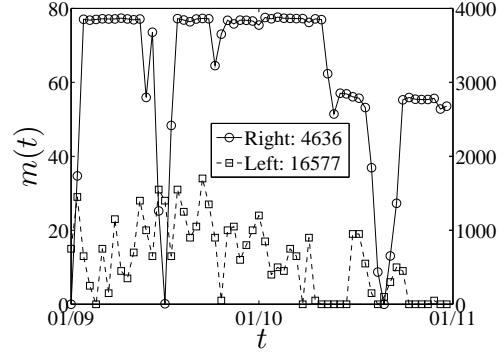}
\caption{\label{Fig:VW:mt} Evolution of the daily number of game
sessions played by two typical avatars 4636 (right axis) and 16577
(left axis).}
\end{figure}

The very small value of $\alpha_m$ indicates that the decay of the
distribution is very slow. The daily online number is no more than
10 for 97.1\% of the avatars and is no more than 1 for 85.2\% of the
avatars. In addition, we notice that the fluctuation at the tail of
the distribution $p(m)$ is high and the occurrence of large $m$
values seems to be greater than the prediction of the $p(m)$
function. The maximal value of the $m$ sequence is 187812 (Avatar
ID: 4636), which means that the avatar went online and offline 128.3
times per hour! The evolution of the daily number $m(t)$ of game
sessions played by this avatar 4636 is illustrated in the right axis
of fig.~\ref{Fig:VW:mt}. We also show in the left axis the evolution
of daily number $m(t)$ of game sessions for avatar 16577 for
comparison.

\begin{figure}[htp]
\centering
\includegraphics[width=6.5cm]{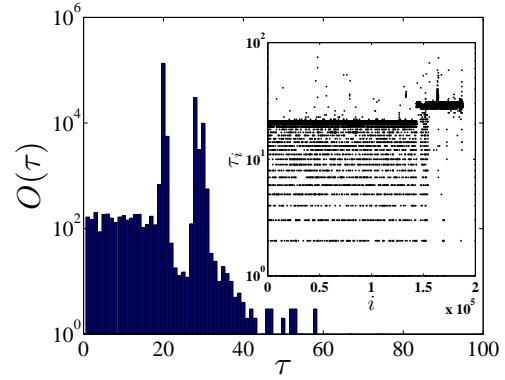}
\caption{\label{Fig:VW:O:tau:ID4636} Occurrence $O(\tau)$ of the
online duration $\tau$ for avatar 4636. The inset shows the
associated $\tau$ sequence.}
\end{figure}

Figure \ref{Fig:VW:O:tau:ID4636} shows the occurrence $O(\tau)$ of
the online duration $\tau$ for avatar 4636. There are two spikes in
fig.~\ref{Fig:VW:O:tau:ID4636} located at around $\tau=20$ and 28.
We observe that $O(20)=134544$ and $O(28)=30302$, which amounts to
71.6\% and 16.1\% of the number of online durations. The inset shows
the associated $\tau$ sequence. A clear change of cheating behavior
from $\tau\approx 20$ to $\tau\approx 28$ is observed, which
happened on 14 October 2007.

\subsection{Distribution of the total time spent by
individual avatars}

An important measure of the avatar game-playing behavior is the
total time he/she spends, which can be calculated as follows,
\begin{equation}
 T_j = \sum_{i=1}^{m_j} \tau_i,
 \label{Eq:Tj}
\end{equation}
which is the sum of all session durations of avatar $j$. The size of
the $T_j$ series is 19843. The maximal total time is 1142 hours
(Avatar ID: 4636), which means that the avatar was active in the
game 18.7 hours per day.

\begin{figure}[htp]
\centering
\includegraphics[width=6.5cm]{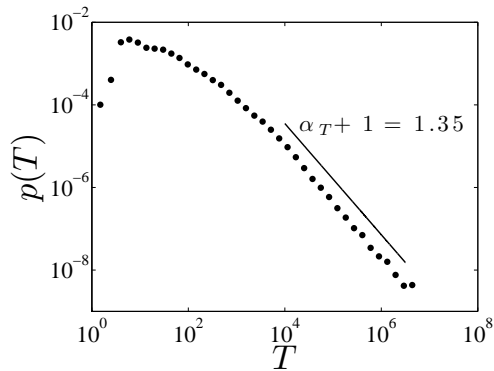}
\caption{\label{Fig:VW:PDF:T} Empirical probability density function
$p(T)$ of the total game-playing time $T$ of 19843 individual
avatars.}
\end{figure}

Figure~\ref{Fig:VW:PDF:T} depicts the probability density function
$p(T)$ of the total time $T$ for the whole population. One can
observe that there is a power-law behavior in the tail of $p(T)$:
\begin{equation}
 p(T)\approx T^{-(\alpha_T+1)}, ~~{\rm{for}}~~T\geqslant{T_{\min}}.
 \label{Eq:pdf:T}
\end{equation}
The tail exponent $\alpha_T$ can also be determined by maximal
likelihood estimation using Eq.~(\ref{Eq:gamma}), where the argument
$m$ is replaced by $T$. By setting $T_{\min} = 500$,
Eq.~(\ref{Eq:gamma}) gives that the tail exponent $\alpha_T = 0.35$.
It is interesting to note that the tail exponent $\alpha_T$ of the
total time $T_j$ is very close to the power-law exponent of the
session number $m_j$.

\subsection{Distribution of the durations of
individual sessions}

\begin{figure}[htb]
\centering
\includegraphics[width=6.5cm]{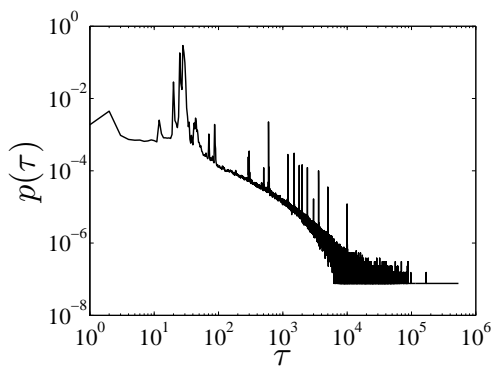}
\caption{\label{Fig:VW:PDF:All:Tau} Empirical distribution density
$p(\tau)$ of the online durations $\tau$. The spikes locate at $\tau
=$ 2, 12, 20, 25, 28, 44, 71, 87, 300, 505, 600, 614, 1200, 1500,
1800, 2000, 2411, 3000, 3600, 5000, and 10000.}
\end{figure}

We put all the online durations $\tau_i$ of all the avatars together
as a whole sample and investigate its distribution. The size of the
whole sample is 13,092,371. Figure \ref{Fig:VW:PDF:All:Tau} shows
the empirical distribution density $f(\tau)$ of the online durations
$\tau$ in log-log scales. The most striking feature of fig.
\ref{Fig:VW:PDF:All:Tau} is the occurrence of many spikes, which
locate at $\tau =$ 2, 12, 20, 25, 28, 44, 71, 87, 300, 505, 600,
614, 1200, 1500, 1800, 2000, 2411, 3000, 3600, 5000, and 10000.
These spikes are outliers that are markedly greater than the normal
level. For some of the spikes, its neighbors are also greater than
the normal level. These spikes indicate the abnormal behavior of
some players, which are usually related to game cheaters. This
observation can be used to identify game cheaters.

\begin{figure}[htb]
\centering
\includegraphics[width=6.5cm]{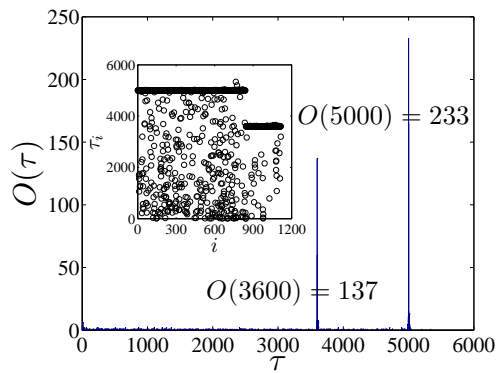}
\caption{\label{Fig:VW:tau:5000} Occurrence $O(\tau)$ of the online
duration $\tau$ for avatar 15611. The inset shows the associated
$\tau$ sequence.}
\end{figure}

Consider the spike at $\tau=5000$. There are 466 game sessions with
$\tau=5000$. We find that there are 15 avatars (IDs: 339, 3797,
5542, 5954, 6418, 6886, 7044, 7767, 10217, 11436, 15611, 15613,
17733, 18075, 18246) whose online duration sequences have at least
one point being $\tau=5000$. The occurrence of $\tau=5000$ is 1 for
all the avatars except for avatars 15611 and 15613, whose
occurrences are 233 and 220, respectively. Figure
\ref{Fig:VW:tau:5000} shows the occurrence $O(\tau)$ of the online
duration $\tau$ for avatar 15611. We find that there are two spikes
in fig.~\ref{Fig:VW:tau:5000} located at $\tau=3600$ and 5000, whose
occurrences are $O(\tau)=137$ and 233. We also observe that
$O(3601)=60$ and $O(5001)=165$. Note that the size of the online
duration sequence of this avatar is 1115. Hence the proportion of
the occurrence of these four $\tau$ values is 53.36\%. The inset
shows the associated $\tau$ sequence. A clear change of cheating
behavior from $\tau=5000$ to $\tau=3600$ is observed, which happend
on 10 October 2007. For avatar 15613, very similar behavior is
observed and a change of cheating from $\tau=5000$ to $\tau=3600$
happened on 12 October 2007. The striking similarity of the behavior
of the two avatars implies that their host players might be closely
related.

\section{Online duration distributions for individual avatars}

Now we turn to study the online-offline behaviors of individual
avatars, which are of potential interest and ultra importance in the
identification of game cheaters, the detection of server traffic,
the understanding of the game-playing patterns of players, and the
design and improvement of online games.

Owning to the consideration of commercial applications and
statistics of the results, we are more interested in active avatars
when investigating their game-playing patterns at the level of
individual avatars. There are numerous avatars whose total numbers
$m$ of online sessions are small. For instance, the proportions of
avatars with $m\leqslant1$, $m\leqslant2$,  $m\leqslant10$,
$m\leqslant50$ and $m\leqslant100$ are 27.8\%, 43.2\%, 66.4\%,
83.6\% and 88.9\%, respectively. Although an avatar with $m=50$ is
not inactive, it is hard to construct its empirical distribution
$p(\tau)$ with sufficient statistics. In addition, according to the
7th Online Game Research Report (2007) and the 8th Online Game
Industry Research Report (2008)\footnote{http://china.17173.com (in
Chinese), accessed on 21 July 2009.}, about 92\% players spent more
than one hour in playing online games every day. Combining these two
facts, we exclude from our analysis the avatars who were online for
no more than 30 days or whose daily cumulative online durations were
less than half an hour. This results in 947 avatars remaining.

As shown in the previous section, especially in
fig.~\ref{Fig:VW:O:tau:ID4636} and fig.~\ref{Fig:VW:tau:5000}, there
are bursts or pulses in the histogram of the occurrence of some
fixed online durations $\tau$. These avatars are impossible to be
operated by humans, rather, they are controlled by some robots,
whose host players are game cheaters. According to the regular
behavior of the program-controlled avatars, we filter out 258 robot
avatars that were too active from the entire population. Finally,
there are 689 avatars remaining for further analysis.

\subsection{Weibull distributions}

\begin{figure}[htp]
\centering
\includegraphics[width=6.5cm]{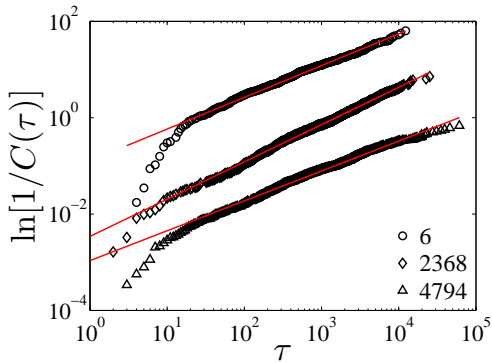}
\caption{\label{Fig:VW:WBL:lnCt} Dependence of
$\ln\left[1/C(\tau)\right]$ as a power-law function of $\tau$ for
three typical avatars (IDs: 6, 2368, 4794).}
\end{figure}

In order to check if these active avatars share the same
online-offline behavior, we determine the empirical complementary
cumulative distribution $C(\tau)$ of each avatar. Our eye-balling
gives us the impression that most distributions have fat tails,
which could be modeled by the Weibull distribution
\cite{Laherrere-Sornette-1998-EPJB,Sornette-2004}
\begin{equation}
C(\tau) = \exp\left[-(\tau/\tau_0)^{b}\right],
 \label{Eq:WBL:Ct}
\end{equation}
where $\tau_0$ is the characteristic time, and $b<1$ is the
exponent. It follows immediately that
\begin{equation}
 \ln\left[1/C(\tau)\right] = (\tau/\tau_0)^{b},
 \label{Eq:WBL:lnCt}
\end{equation}
which means that $\ln\left[1/C(\tau)\right]$ scales as a power law
with respect to $\tau$. Figure \ref{Fig:VW:WBL:lnCt} shows the
dependence of $\ln\left[1/C(\tau)\right]$ as a function of $\tau$
for three avatars. All the three curves exhibit power laws with the
scaling ranges spanning about three orders of magnitude, which is
the graphic evidence that the distribution of the online durations
for individual avatars of this type is Weibull.

In order to identify the avatars whose online durations conform to
the Weibull distribution, we design an approach to classify the
avatars based on statistical tests. For each avatar, its empirical
distribution of online durations is fitted to a Weibull formula by
means of the maximum likelihood estimation (MLE) method. The fitted
formula is then converted to its cumulative form $F(\tau)$. We then
investigate whether the sample of online durations is drawn from the
``theoretical'' distribution $F(\tau)$ from the best MLE fit. The
null model is that the data can be modeled by a Weibull
distribution. We can perform the Kolmogorov-Smirnov (KS) test
\cite{Smirnov-1948-AMS,Young-1977-JHcCc} for this purpose. The
Kolmogorov-Smirnov statistic (KS statistic), which measures the
distance between the empirical cumulative distribution function of
the sample and the cumulative distribution function of the best fit,
is defined as
\begin{equation}
{\rm{KS}} = \max(|F_{\rm{emp}} - F|),
 \label{Eq:KS}
\end{equation}
where $F_{\rm{emp}}$ is the cumulative distribution function of the
empirical sample and $F$ is the cumulative distribution function
from the best MLE fit. Alternatively, the Cram\'{e}r-von Mises
criterion can also be used for judging the goodness-of-fit of the
probability distribution compared with a given
distribution~\cite{Darling-1957-AMS}, which is given by
\begin{equation}
C_{M}^{2}=n\int_{-\infty}^{+\infty}\left[F(\tau)-F^{*}(\tau)\right]^{2}dF(\tau).
 \label{Eq:CvM1}
\end{equation}
In one-sample applications, the function can be described as
follows~\cite{Pearson-Stephens-1962-Bm,Stephens-1970-JRSSB},
\begin{equation}
C_{M}^{2}=\frac{1}{12n}+\sum_{i=1}^{n}\left[\frac{2i-1}{2n}-F(\tau_{i})\right]^{2}~,
 \label{Eq:CvM2}
\end{equation}
where $n$ is the sample size. If the KS (or CvM) statistic is less
than a critical value, the null hypothesis cannot be rejected.

\begin{figure}[htp]
\centering
\includegraphics[width=6.5cm]{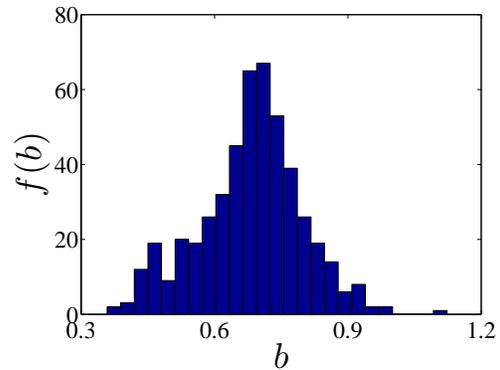}
\caption{\label{Fig:VW:WBL:Hist:b} Histogram of the fitted exponent
$b$ for the 489 avatars.}
\end{figure}

At the significant level of 1\%, we find that there are 489 avatars
whose online durations can be well modeled by the Weibull
distribution. Figure \ref{Fig:VW:WBL:Hist:b} presents the histogram
of the fitted exponent $b$ for the 489 avatars. There is one value
of $b$ (ID: 5483) that is greater than 1, which corresponds to a
sub-exponential distribution decaying faster than exponential. We
find that the distribution is mono-modal and $b=0.68\pm0.12$.

\subsection{Other distributions}

\begin{figure}[htp]
\centering
\includegraphics[width=6.5cm]{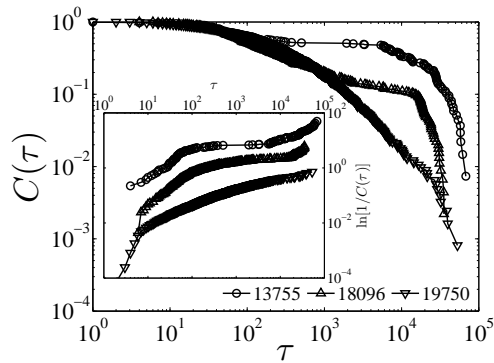}
\caption{\label{Fig:VW:PDF:other} Survival distributions $C(\tau)$
of the online durations $\tau$ for three typical avatars (IDs:
13755, 18096, 19750). The inset shows correspondingly the plots of
$\ln\left[1/C(\tau)\right]$ versus $\tau$.}
\end{figure}

For the avatars whose online durations do not follow Weibull
distributions, we cannot find a simple form for the online duration
distribution. Figure \ref{Fig:VW:PDF:other} illustrates the survival
distributions of $\tau$ for three typical avatars in log-log scales.
It seems that the first-order derivative is not continuous for
avatars 13755 and 18096, since there are clear kinks in the
$C(\tau)$ curves. For avatar 19750, the $C(t)$ curve looks like a
Weibull truncated with a power-law tail. However, statistical tests
shows that it is neither a Weibull distribution nor a power-law
tailed distribution. The inset of fig.~\ref{Fig:VW:PDF:other} shows
correspondingly the curves of $\ln\left[1/C(\tau)\right]$ with
respect to $\tau$ for the three avatars. No evident power-law regime
is observed in the three curves, which confirms that the online
durations of these avatars do not follow Weibull distributions.

\section{Conclusion}

In summary, we have studies the online-offline activities and
game-playing behaviors of avatars in a massive multiplayer online
role-playing game based on the log files recorded during the time
period from 1 September 2007 to 31 October 2007. We found that the
number of game sessions and total time of online durations of
individual avatars are distributed according to a power law, with
large bursts in both tails. In addition, the distribution of the
online durations of all avatars as a whole sample is decorated by
sharp spikes. These phenomena are signals of game cheaters who used
robots to control their avatars, which can be identified by the
abnormal pulses in the distribution of online durations for
individual avatars. In addition, we also found that there are a
group of normal avatars whose online durations are distributed as
Weibulls. These findings have potential applications in the online
game industry.

Our finding that the online durations of many {\em{normal}} avatars
are distributed according to a Weibull distribution adds new
evidence that human dynamics is not a simple Poisson process
\cite{Barabasi-2005-Nature}. However, the Weibull behavior cannot be
explained by existing models based on priority queue
\cite{Barabasi-2005-Nature}, cascading nonhomogeneous Poisson
process \cite{Malmgren-Stouffer-Motter-Amaral-2008-PNAS}, or
adaptive interest \cite{Han-Zhou-Wang-2008-NJP}.

\acknowledgments{This work was partly supported by the Program for
New Century Excellent Talents in University (Grant No.
NCET-07-0288).}

\bibliographystyle{eplbib} 
\bibliography{E:/papers/Auxiliary/Bibliography}

\end{document}